\pdfoutput=1
\documentclass[conference,a4paper]{IEEEtran}

\usepackage{amsmath}
\ifCLASSINFOpdf
   \usepackage[pdftex]{graphicx}
\else
\fi
\begin{document}
\title{The Interplay of European Hydro Power and North African Solar Power in a Fully Renewable European Power System }

\author{\IEEEauthorblockN{Alexander Kies, Bruno Schyska, Lueder von Bremen}
\IEEEauthorblockA{ForWind - Center for Wind Energy Research\\
University of Oldenburg
Oldenburg, Germany\\
Email: alexander.kies@uni-oldenburg.de }}%\and
%\IEEEauthorblockN{Bruno Schyska, Lueder von Bremen}
%\IEEEauthorblockA{ForWind, Center for Wind Energy Research\\
%Oldenburg University\\
%26129 Oldenburg, Germany}\and
%\IEEEauthorblockN{Martin Greiner}
%\IEEEauthorblockA{Department of Engineering, Aarhus University\\
%8000 Aarhus C, Denmark}}


%Telephone: (800) 555--1212, Fax: (888) 555--1212}


\maketitle

\begin{abstract}
A fully renewable European power system comes with a variety of problems. Most of them are linked to the intermittent nature of renewable generation from the sources of wind and photovoltaics. 
A possible solution to balance European generation and consumption are European hydro power with its seasonal and North African Concentrated Solar Power with its daily storage characteristics. \\
In this paper, we investigate the interplay of hydro and CSP imports in a highly renewable European power system.
We use a large weather database and historical load data to model the interplay of renewable generation, consumption and imports for Europe.\\
We introduce and compare different hydro usage strategies and show that hydro and CSP imports must serve different purposes to maximise benefits for the total system. 
CSP imports should be used to cover daily deficits, whereas hydro power can cover seasonal imbalances.
If hydro is used in a ``Hydro First'' strategy, only around one quarter of North African Solar Power could be exported to Europe, whereas
this number increases to around 60\%, if a cooperative hydro strategy is used.
\end{abstract}



\section{Introduction}
Renewable power sources can solve the major problems associated with conventional fossil and nuclear generation:
greenhouse gas emissions, deplenishing of fossil resources and waste generation.
The European power system has already transformed to some extent and obtains around one quarter of its electricity consumption from renewable sources.
In addition, wind and photovoltaics (PV) shares are expected to continue growing (\cite{Eurostat_2015}).
However, generation from wind and PV has intermittent patterns, which makes their system integration difficult. 
\\ Several research papers have adressed the difficulties of intermittent generation for the European power system and investigated possible solutions;
among them are the influence of excess generation (\cite{Heide_2011}), optimising the mix of renewable generation from different sources like wind, PV, and wave (\cite{Heide_2010}, \cite{Kies_2015}, \cite{Kies_2016wave}, \cite{franccois2016increasing}, \cite{jurasz2017integrating}, \cite{jurasz2017modeling}), 
storage and balancing technologies (\cite{Rasmussen_2012}, \cite{Weitemeyer_2015}, \cite{weitemeyer2015european}, \cite{schlachtberger2016backup}) or transmission grid extensions \cite{Becker_2014}, \cite{Steinke_2010}, \cite{kies2016curtailment}).
Other possible options include the modification of the demand side (\cite{kleinhans2014towards}, \cite{Kies_2016dsm}) or system-friendly renewables (\cite{hirth2016system}, \cite{chattopadhyay2017impact}).
\\Concentrated solar power (CSP) is a renewable energy technology that produces like PV useable power from solar energy.
CSP has different properties that make it interesting among other renewable power sources.
It converts concentrated light to heat, which can either be stored or used to drive a heat engine connected to an electrical power generator.
Thermal energy can be easier stored than electrical energy, which is one major advantage of CSP.
However, CSP has a higher levelised cost of electricity than PV and this is not expected to change in the near future (\cite{FraunhoferISE_2013}).
\\ The idea to couple the European power system with solar power imports from North Africa was 
already proposed at multiple opportunities: The most famous one was the DESERTEC project; a large scale project to implement a global renewable energy plan 
based on the idea of harvesting renewable resources at suitable sites and transferring the generated power via high voltage connections to Europe.
It focussed on concentrated solar power, because CSP technologies are able to provide dispatchable power to compensate variable generation
from wind and photovoltaics (PV). \\
The technical potential of solar power generation in the Middle East and North Africa (MENA) region was estimated by Trieb et al. (\cite{trieb2012solar}) to be 538,000 TWh per year.
It was proposed that by 2050 roughly 15\% (700 TWh/a) of the European demand could be covered by CSP from the MENA region.
This would require high voltace direct current (HVDC) transmission capacities of 117 GW between Europe and NA and investments of approximately 700 billion Euro at prices of 2010 (85\% CSP infrastructure, 15\% transmission links).
117 GW of capacity and 700 TWh/a of transferred energy correspond to an average usage rate of 70\%, which is significantly higher than the values obtained in this investigation. The differences are likely caused by a different methodology. The potential of CSP to balance and support a European power system was further investigated 
by Viebahn, et. al. and Trieb et al.
(\cite{Viebahn_2011},
\cite{trieb2006trans}).\\
Today, there are different concepts for CSP generation plants to harvest solar energy.
Among them are parabolic trough, parabolic dishes, linear Fresnel reflectors and towers (\cite{Guerrero_2013}).
Usually a DNI of around 1800 kWh/m$^2$ is considered to be technically required for csp power plants to be operated.
This is given in some parts of Southern Europe. 
However, North Africa reaches DNI values of 2,500 kWh/m$^2$ and more.
Hence, North African countries have an outstanding potential for the use of solar power and the intermittency of excess CSP generation
could be further used for 
specific applications such as seawater desalination.

\section{Data and Methodology}
  \begin{figure}
\centering
\includegraphics[width=0.5\textwidth]{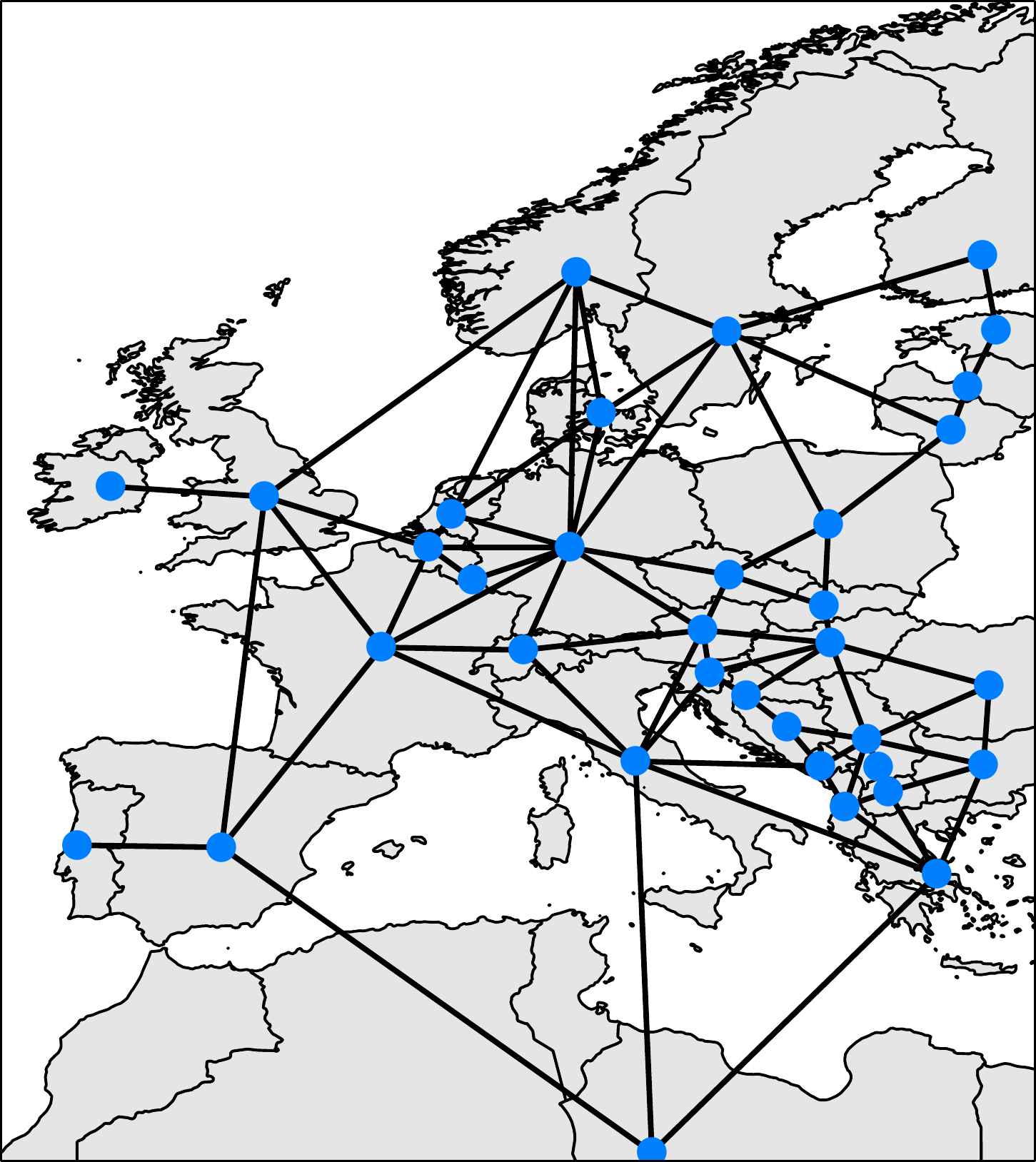}
\caption{Topology of the investigated power system. Nodes are represented by dots, which are connected by inter-connecting transmission links.}
\label{fig:topo}
\end{figure}
  \begin{figure}
\centering
\includegraphics[width=0.5\textwidth]{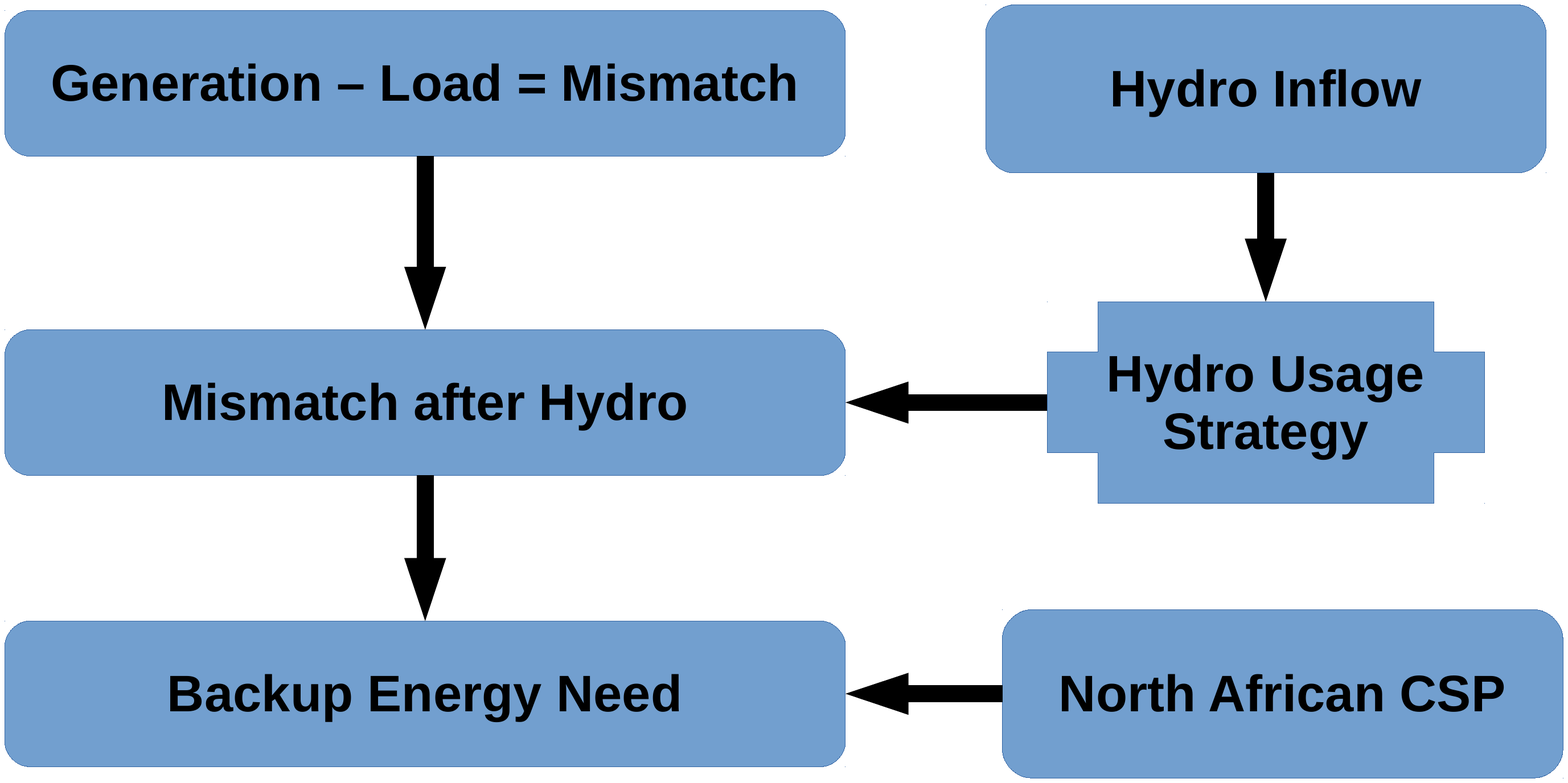}
\caption{Flowchart of the calculations in this paper. The hydro usage strategy is varied.}
\label{fig:flowchart}
\end{figure}
 A fully renewable European power system still requires an enormous amount of backup energy to be stable (\cite{Heide_2010}).
  An idea is to support the European power system via imports of solar power from Northern Africa.
In principle, it is possible and desirable that also the option of exporting energy from Europe to North Africa at times of generation surplusses is considered. However, we neglect this aspect in this work, 
and focus on imports only:
If there is demand from Europe and possible supply from North Africa and transmission capacities are available, then power is transferred.
Furthermore, no assumption on the market mechanism behind this exchange is made. 
The topology of the simplified power system is shown in Fig. \ref{fig:topo} and the methodology behind the calculations in this paper in Fig. \ref{fig:flowchart}.
We have calculated generation from the renewable sources of wind and photovoltaics (PV) as well as energy inflow into hydro storage for Europe and into CSP storage for North Africa. 
All European countries are assumed to be interconnected with unlimited transmission capacities (copperplate approximation).
Furthermore, the European power system is connected to North-Africa via a link that allows power transmission into the direction North Africa $\rightarrow$ Europe.
\subsection{Model}
For every node, we have computed a generation time series
\begin{align*}
 G_n(t) &= W_n(t) + S_n(t),
\end{align*}
consisting of the single generation time series for wind $W(t)$ and PV $S(t)$. 
Together with the time series of the load $L_n(t)$, the time series of the (generation-load-)mismatch was directly computed as
\begin{align}
 \Delta_n(t) &= G_n(t) - L_n(t).
\end{align}
Because of the copperplate approximation and since we are not interested in transmission within Europe, we aggregate the mismatches of the single countries to one European mismatch time series,
\begin{align}
 \Delta (t) &= \sum_n \Delta_n(t).
\end{align}
For stable operation of the power system, it is necessary that generation and consumption are balanced, i.e., the mismatch needs to vanish.
This results in the balancing equation,
\begin{align}
\Delta (t) &= C(t) - G^H(t) - I(t) - B(t),
\label{eq:balancing}
\end{align}
where $C(t)$ is the curtailment of renewable power (wind/PV), $B(t)$ is the backup generation provided from dispatchable conventional sources,
$G^H(t)$ is generation from hydro and $I(t)$ are imports from North Africa.
The left part of the balancing equation is the active part that is determined by the given data while the right side is the reactive part,
The ratio of average generation to average load in a given country $n$ is described by the share of renewables $\alpha_n$, defined by
\begin{align}
 \alpha_n &= \frac{\left<G_n(t) + G^H_n(t)\right>}{\left<L_n(t)\right>}. \label{eq:alpha}
\end{align}
Throughout this paper, $\alpha_n$ is set to unity for all countries.
Beside $\alpha$, the solar share $\beta_n$ is of importance, defined via
\begin{align*}
 \beta_n &= \frac{\left<G^S(t)\right>}{\left<G^S(t) + G^W(t)\right>},
\end{align*}
and adopted from scenario B published by Fraunhofer ISI in 2011 (\cite{pfluger2011tangible}) for all European countries.
From Eq. \ref{eq:balancing}, we compute the time series of the balancing via
 \begin{align*}
  B(t) &= -\min\{ \Delta(t)+ G^H(t) + I(t), 0\},
 \end{align*}
 and the curtailment via
 \begin{align*}
  C(t) &= \max\{ \Delta(t), 0\}.
  \end{align*}
The need for backup energy in a given time interval is consequently defined as
\begin{align*}
 E^B = \int_T B(t) dt
\end{align*}
and the need for backup capacity via
\begin{align}
 0.99 = \int_0^{B^C} p(B(t)) dp,
 \label{eq:backupcap}
\end{align}
where $p(.)$ is the distribution function of backup events.
Hence, the backup capacity is via definition sufficient in 99\% of the time. This value below 100\% is chosen to reduce the sensitivity towards the weather database. 

\subsection{Generation and Load Data}
We used ten years of weather data
with a spatial resolution of $7 \times 7$ km and an hourly temporal resolution to model generation from wind and PV.
Generation is first computed for each grid cell and in a second step aggregated to the country level.
The weather data consists of downscaled windspeeds and 2 m temperatures from the MERRA reanalysis (\cite{rienecker2011merra})
and irradiation from satellite images (Meteosat First Generation (MFG) and Meteosat Second Generation (MSG)).
To calculate feed-in from wind, we used the
power curve of an Enercon E-126 at $140$ m hub height with $5$\% plain losses for every grid cell. \\
Surface irradiances for PV power simulations were obtained using the Heliosat method (\cite{Cano_1986}, \cite{Hammer_1998}).
The conversion of global horizontal irradiance to irradiation on the tilted modules was based on the Klucher model (\cite{Klucher_1979}).
Finally,
DC power was converted into AC power using the parameter of a Sunny Mini Central 8000TL converter.
To model the distribution of wind and PV generation facilities within each country, the German distribution of capacities for wind and PV in dependency of the available resource (average wind speed / average global horizontal irradiation) was empirically derived
and adopted for every country. 
The mix of installed capacities of wind and PV was adopted for every country from a scenario by Fraunhofer ISI (\cite{pfluger2011tangible}).
\\To model hydro power, we chose a potential energy approach using runoff data from the ERA-Interim reanalysis (\cite{dee2011era}).
The potential gravitational energy of a mass $m$ relative to the sea level is given by
\begin{align*}
 U &= m g h, 
\end{align*}
where $g = 9.81 \text{m}/\text{s}^2$ is the constant of gravititational acceleration on Earth and h the height above sea level.
Inflow into hydro storages is calculated as a linear function of the potential energy of the runoff
\begin{align*}
 I^H_n(t) &=  f \int_A g m(t) h(x,y) dA.
\end{align*}
$f$ is a normalization constant to enforce $\left<I^H_n(t)\right>$ = $\left<G(t)^H_n\right>$, where $G_n^H$ is today's average hourly generation from hydro in the corresponding country.
Thus, we assume that the generation from hydro does not increase significantly in the future.
This seems reasonable because European hydro is already well exploited today.
Data on today's generation was collected from different sources.
The mass is calculated from the runoff data.\\
Irradiation data for CSP calculations was retrieved like irradiance data for PV power calculations from satellite images (MSG).
Thermal energy generated from CSP plants was computed using a parametric model (Solar Advisor Model (SAM)) developed by NREL (\cite{SAM_NREL}).
For the distribution of capacities, a homogeneous distribution over a reduced spatial resolution of 
$0.5^\circ \times 0.5^\circ$ was assumed to be reasonable. \\
More details on generation data are given in the project report by Kies et al. (\cite{restorereport2016}). \\
Load time series were derived from historical load data provided by ENTSO-E.
This load data was further modified within the RESTORE 2050 project to account for expected shares from e-mobility and heat pumps in the future electricity consumption.

\section{Hydro Usage Strategies}
In this paper, we investigate the impact of four different heuristic hydro usage strategies on the interplay of European hydro power and North African CSP.
We believe that these heuristic strategies could be realised by a proper market environment.
In general, generation from hydro power $G_n^H(t)$ is given in our model by
  \begin{align*}
   \hat{G}_n^H(t) &= \max\left(\{0,-\Delta'_n(t)-\lambda\left[S_n^H(t), t\right] \cdot \frac{\left<I_n\right>}{\sum_n \left<I_n\right>}\left<L\right>\}\right), \\
G_n^H(t) &= \tau\left[S_n^H(t), t\right] \cdot \max\left(\{\hat{G}_n^H(t),G^H_{+,n}, S_n^H(t)\}\right), \\
 \end{align*}
 where $\hat{G}_n^H(t)$ is the hypothetical generation without power or energy constraints, 
 $S_n^H(t)$ is the hydro storage filling level and $G^H_{+,n}$ is the installed hydro power capacity.
 The $[..]$ indicate that $\lambda$ and $\tau$ can be a function of a function, e.g. the storage filling level.
 If the generation is known, the storage level of the following time step is given by
 \begin{align*}
  S_n^H(t+1) &= \min\{S_n^H(t) + I_n^H(t) - G_n^H(t),S^H_{+,n}\}, 
 \end{align*}
 where $S^H_{+,n}$ is the storage reservoir capacity.
 We investigate four different hydro strategies within this paper with two goals: i) The reduction of the need for backup capacities in an isolated European power system.
 ii) A better integration of North African CSP in an extended European power system, i.e. the reduction of the need for backup energy.
 For the latter goal we refer to the strategies (ii) - (iv) as \textit{cooperative} strategies, because they allow for a cooperative use of European hydro power and North African CSP.
\subsection{Hydro First}
The first strategy used simply reads
\begin{align*}
 \lambda = 0,\\
 \tau = 1,
\end{align*}
and is entitled as ``Hydro first'',
because it uses hydro power, whenever there is a mismatch after wind and PV without consideration of possible CSP imports.
If there remains a mismatch after the usage of hydro power, CSP power is imported.

\subsection{Fixed Threshold}
The second strategy fixes $\lambda$ at a constant value,
\begin{align*}
 \lambda &= c \in [0,1),\\
  \tau &= 1.
\end{align*}
Hence, hydro power is only used above a fixed threshold to cover residual loads. Anything below that threshold is not balanced via generation from hydro.
For $\lambda=0.1$, for instance, hydro power is dispatched if the mismatch is greater than 10\% of the average load.
\subsection{Variable Threshold}
The third strategy does not assign $\lambda$ a constant value, but instead assumes time dependency,
\begin{align}
 \lambda(t) &= \lambda_i \in [0,1) , t \in T_i, \cup_i T_i = T, \\
 \tau &= 1,
\end{align}
where we assumed $T_i$ to be the months of the year, $T_i \in $[January, ..., December]. This can be interpreted as taking the seasonality of the inflow and the mismatch into account.
The coefficients $\lambda_i$ are determined such that the need for backup capacity is minimised:

\begin{align}
   \underset{\boldsymbol{\lambda_i}}{\text{minimise }}
 \kappa_B^C \\
 \text{subject to } 
 0 \leq \lambda_i < 1 \label{eq:variablethreshold}.
\end{align}

\subsection{Bernoulli}
Finally, we propose a method which considers the storage filling level $S_n^H(t)$. We refer to it
as ``Bernoulli'', because $\tau$ is given as the solution for the flow from a tank through a small orifice via the Bernoulli equation.
In this case the coefficients read

\begin{align*}
\lambda &= c \in [0,1) ,\\
\tau &= \begin{cases}
         \sqrt{\frac{2 S^H_n(t) - \kappa^H_n}{2 \kappa^H_n}} + 1, \text{ if } S^H_n(t) \geq \frac{\kappa^H_n}{2} \\
         1 - \sqrt{\frac{\kappa^H_n - 2 S^H_n(t)}{2 \kappa^H_n}}, \text{ if } S^H_n(t) < \frac{\kappa^H_n}{2}.
       \end{cases}
\end{align*}
$\lambda$ could, instead of a being constant, chosen to be variable.
\section{North African Concentrated Solar Power}
After the use of hydro power, the use of North African CSP is determined. 
This is done according to
\begin{align*}
 I(t) &= \max\{0,-\Delta(t) - G^H(t),S^{NA}(t), G^{NA}_+\},
\end{align*}
where $S^{NA}(t)$ is the storage filling level and $G^{NA}_+$ the installed CSP power,
which can be either be interpreted as the generation capacity of CSP or the transmission capacity between Northern Africa and Europe.
The storage filling level of each step was calculated equivalently to the filling level of the hydro reservoirs via
\begin{align*}
  S^{NA}(t+1) &= \min\{S^{NA}(t) + I^{NA}(t) - I(t),S^{NA}_+\}.
 \end{align*}
 $S^{NA}_+$ is the storage energy capacity of North African CSP. It is, like the generation capacity, varied in the results section.

\section{Results}
 \subsection{European Hydro Power}
\begin{figure}
\centering
\includegraphics[width=0.5\textwidth]{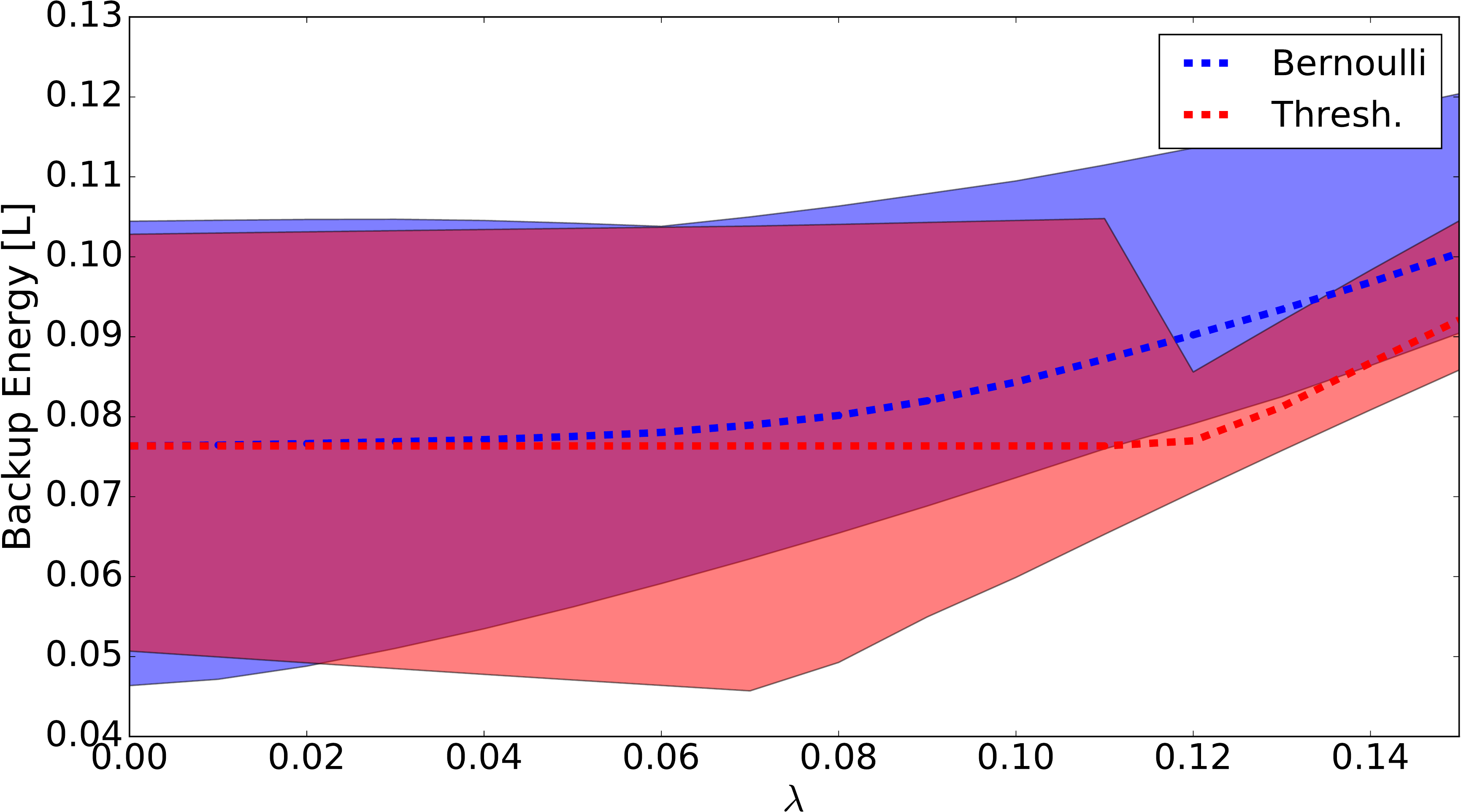}\\\includegraphics[width=0.5\textwidth]{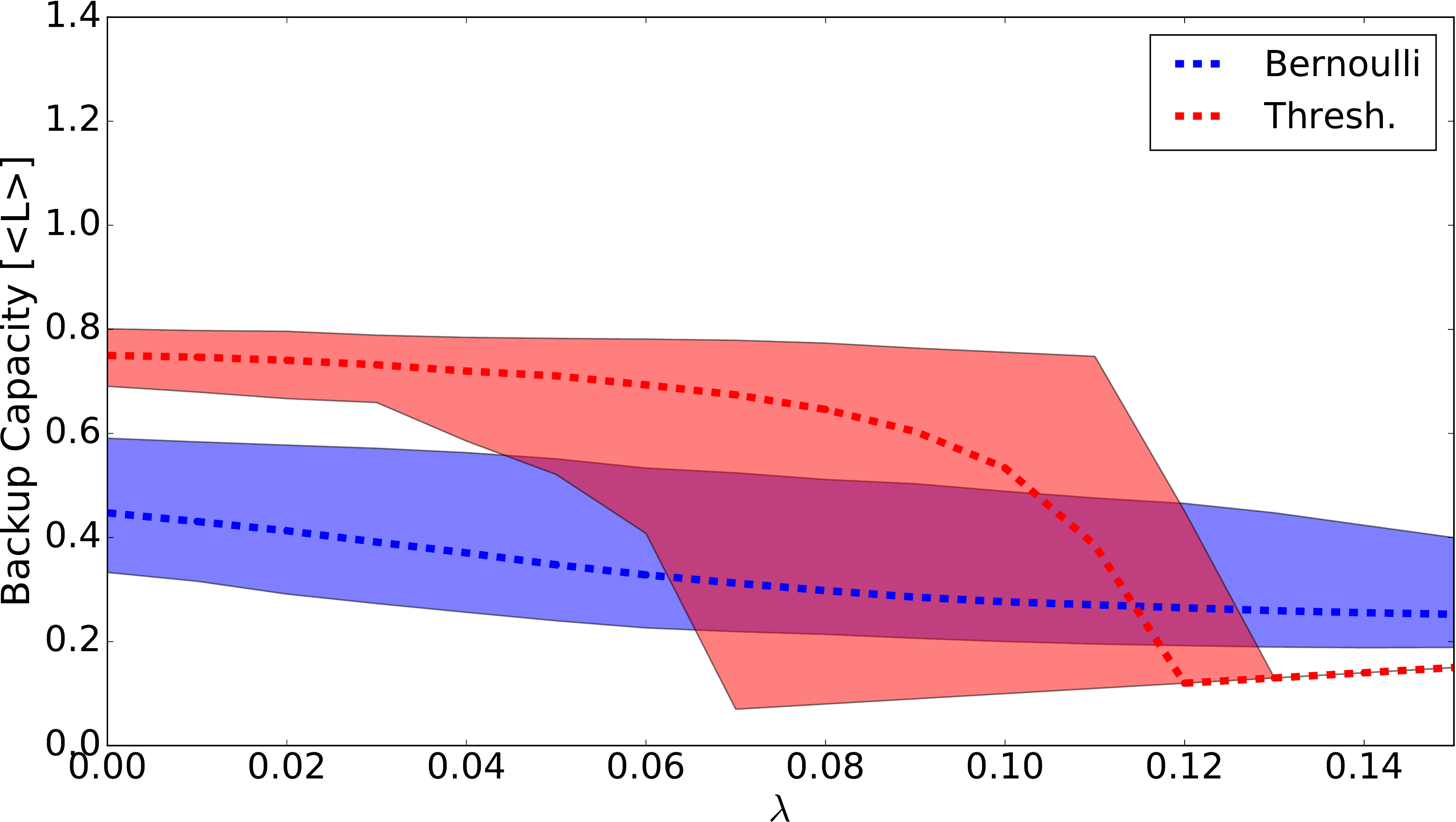}
\caption{European need for backup energy (top) and backup capacity (bottom) for the proposed hydro usage strategies in Europe. Computations were carried out for ten years of generation and load data.
The colored areas indicate the yearly variation. The ``hydro first`` strategy is represented by $\lambda=0$ of the red curves.}
\label{fig:backupebackupc}
\end{figure}

 First, we investigate the effect of the different hydro strategies without imports of North African CSP.
 We started by determining the solution of the optimisation problem for Eq. \ref{eq:variablethreshold}.
 However, the optimal solution turned out to be $\lambda_i = 0.117 \pm 0.02 \forall i$.
 Therefore, we set $\lambda = 0.117$ and skip the distinction between hydro strategies variable and fixed threshold, and refer to it as ``threshold'' in the following.
 Figure \ref{fig:backupebackupc} shows the need for backup energy and backup capacity in dependency of the constant $\lambda$ for 
 all years, where the colored areas indicate the yearly variation.
 The need for backup energy remains practically constant over a wide range of $\lambda$ below 8\% of the consumption in both cases.
 In the case of the Bernoulli strategy the need starts increasing at around $\lambda=0.04$  and in the case of the threshold at around $\lambda=0.12$.
 This can be interpreted as being caused by the rare occurrence of mismatches above 12\% of the average load. Therefore, hydro storages can not exploit their full potential to cover energy deficits, which in turn eventually leads to spillage. 
  For $\lambda=0$ it can be observed that the need for backup capacity is high at ca. 75\% of the average load. 
 This value drops to 10\% of the average hourly load at $\lambda=0.12$ for the threshold strategy.
 For Bernoulli, values start at around 45\% and constantly decrease to 25\% at $\lambda=0.15$.
 Furthermore, it can be observed that the yearly variation of backup capacities is much larger in the case of the Threshold strategy. 
 In the case of the Bernoulli strategy, hydro storages react slower to residual mismatches and yearly variation is therefore smaller.

 \subsection{Hydro Power and CSP}
   \begin{figure}
\centering
\includegraphics[width=0.5\textwidth]{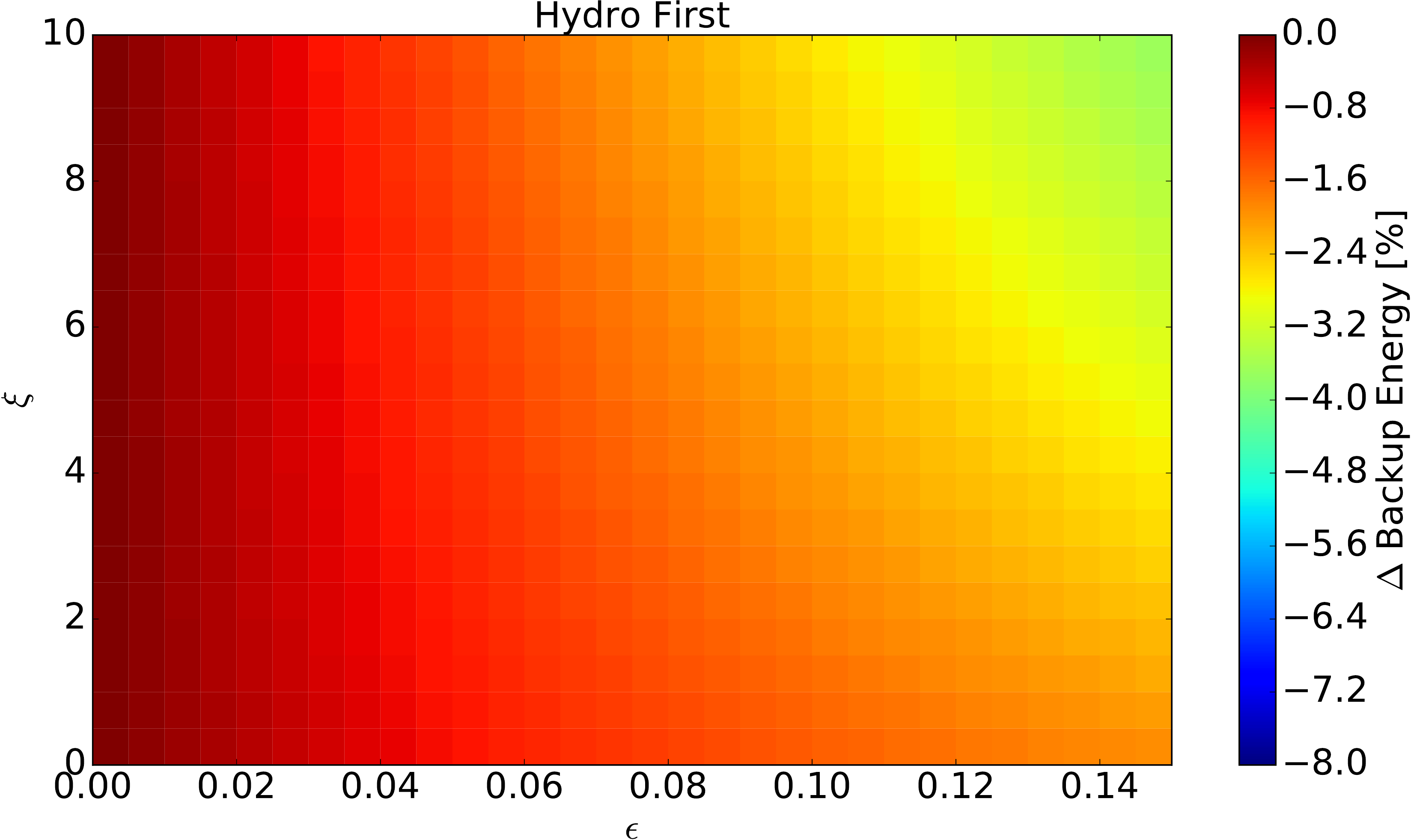}
\\\includegraphics[width=0.5\textwidth]{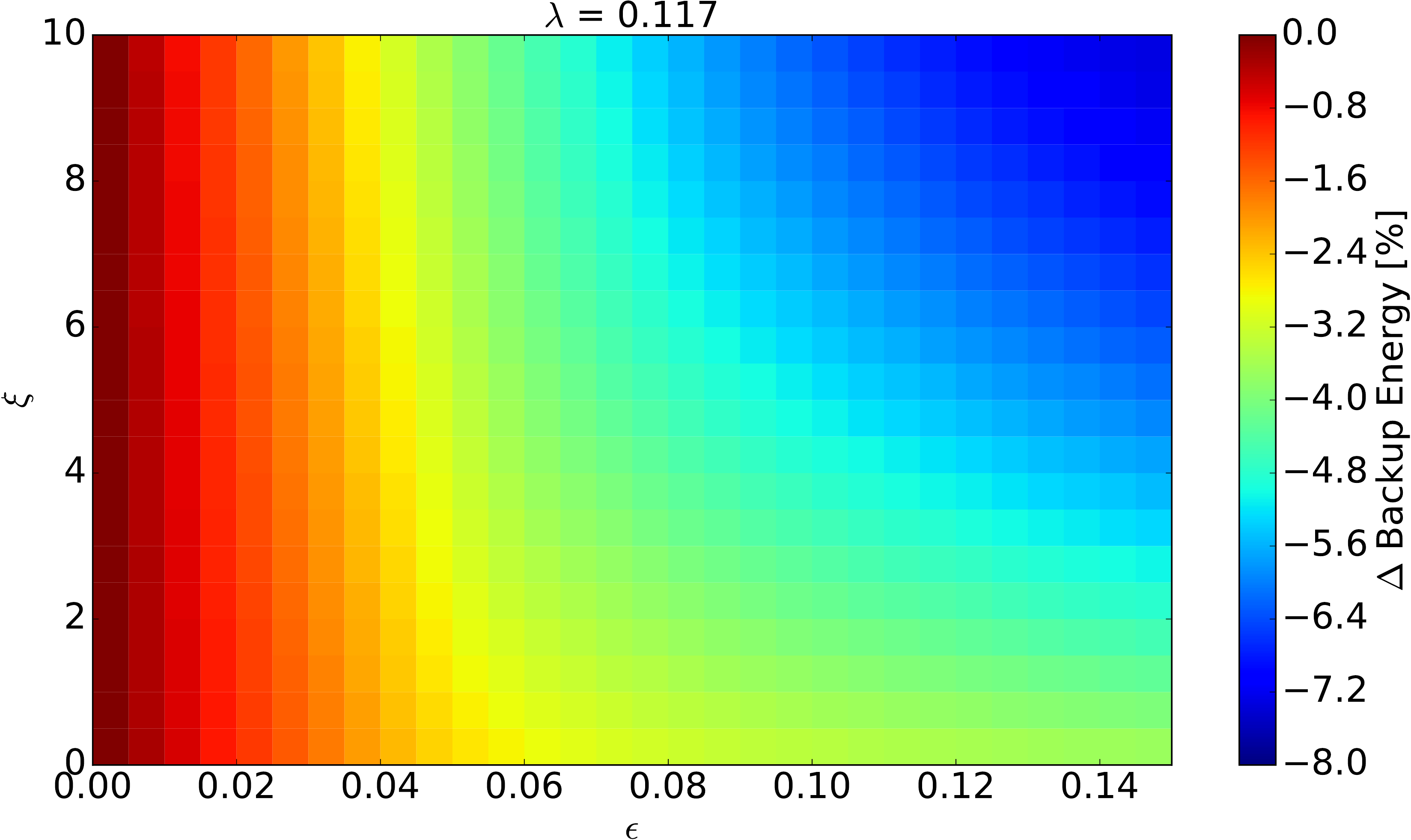}\\
\includegraphics[width=0.5\textwidth]{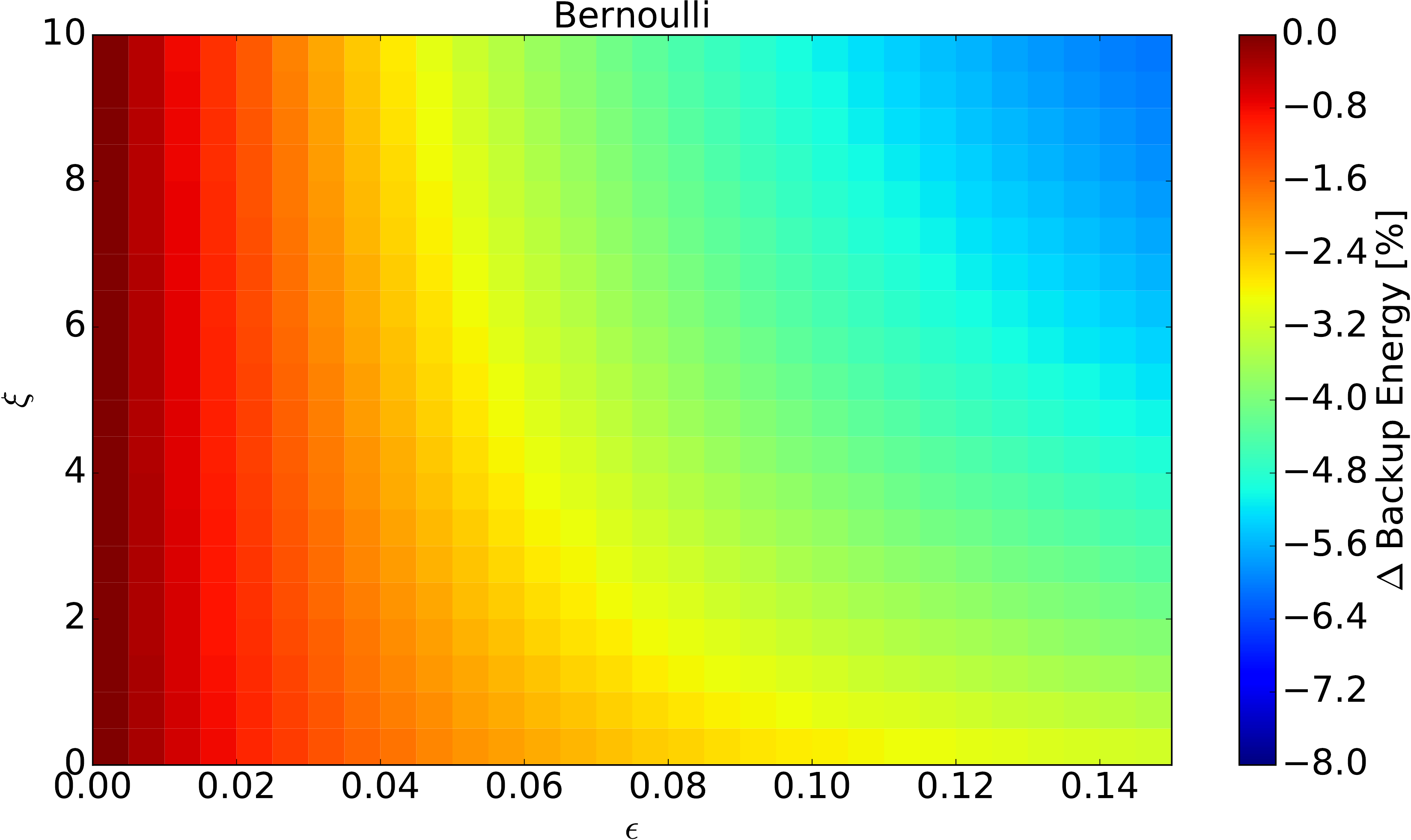}
\caption{Reduction of the European backup energy need using North African solar imports in dependency of the average CSP generation and the CSP storage capacities for three different hydro usage strategies. Top: hydro first, middle: threshold, bottom: Bernoulli. For details, we refer to the text.}
\label{fig:backupebackupc2}
\end{figure}
\begin{figure}
\centering
\includegraphics[width=0.5\textwidth]{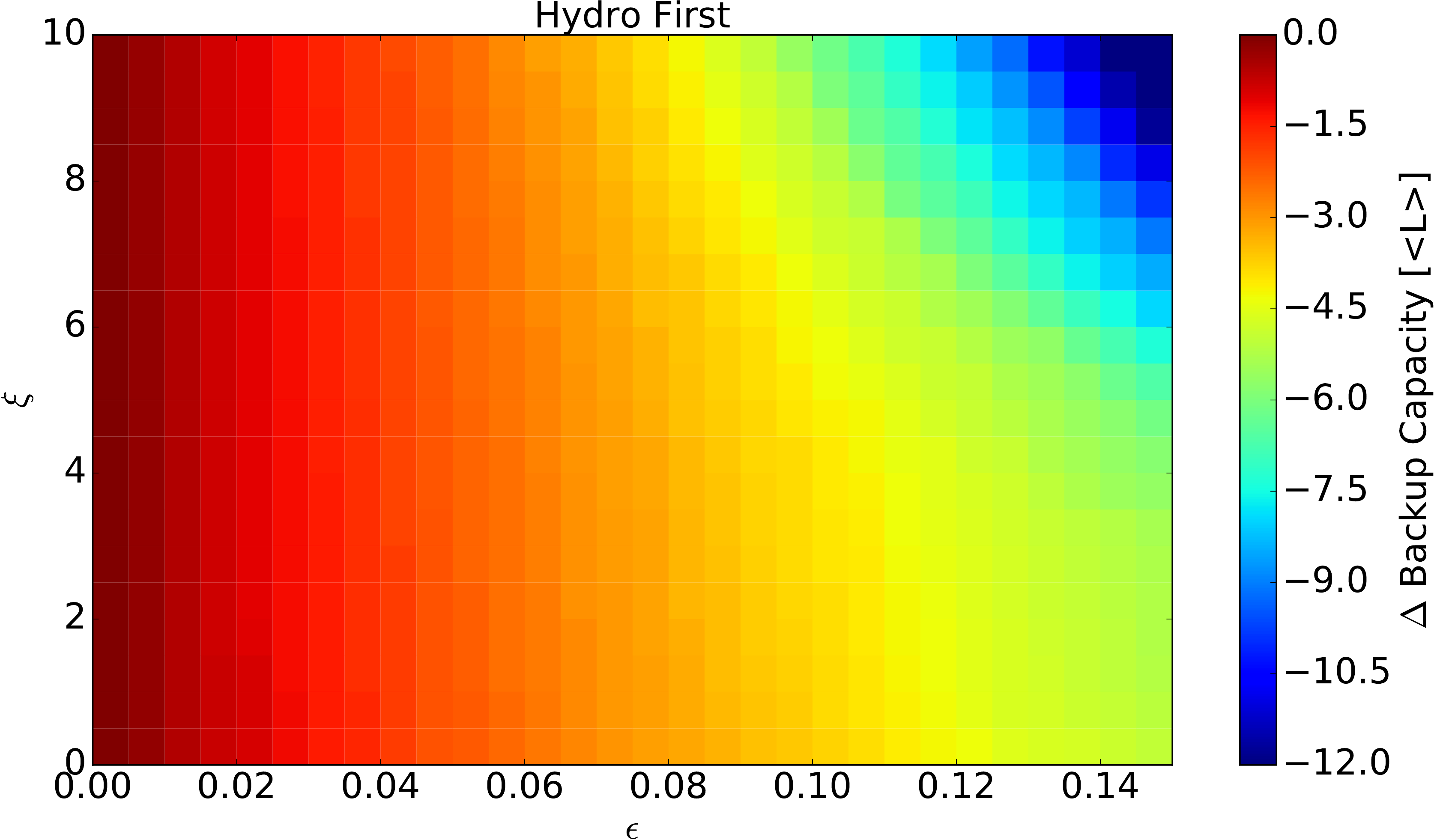}
\\\includegraphics[width=0.5\textwidth]{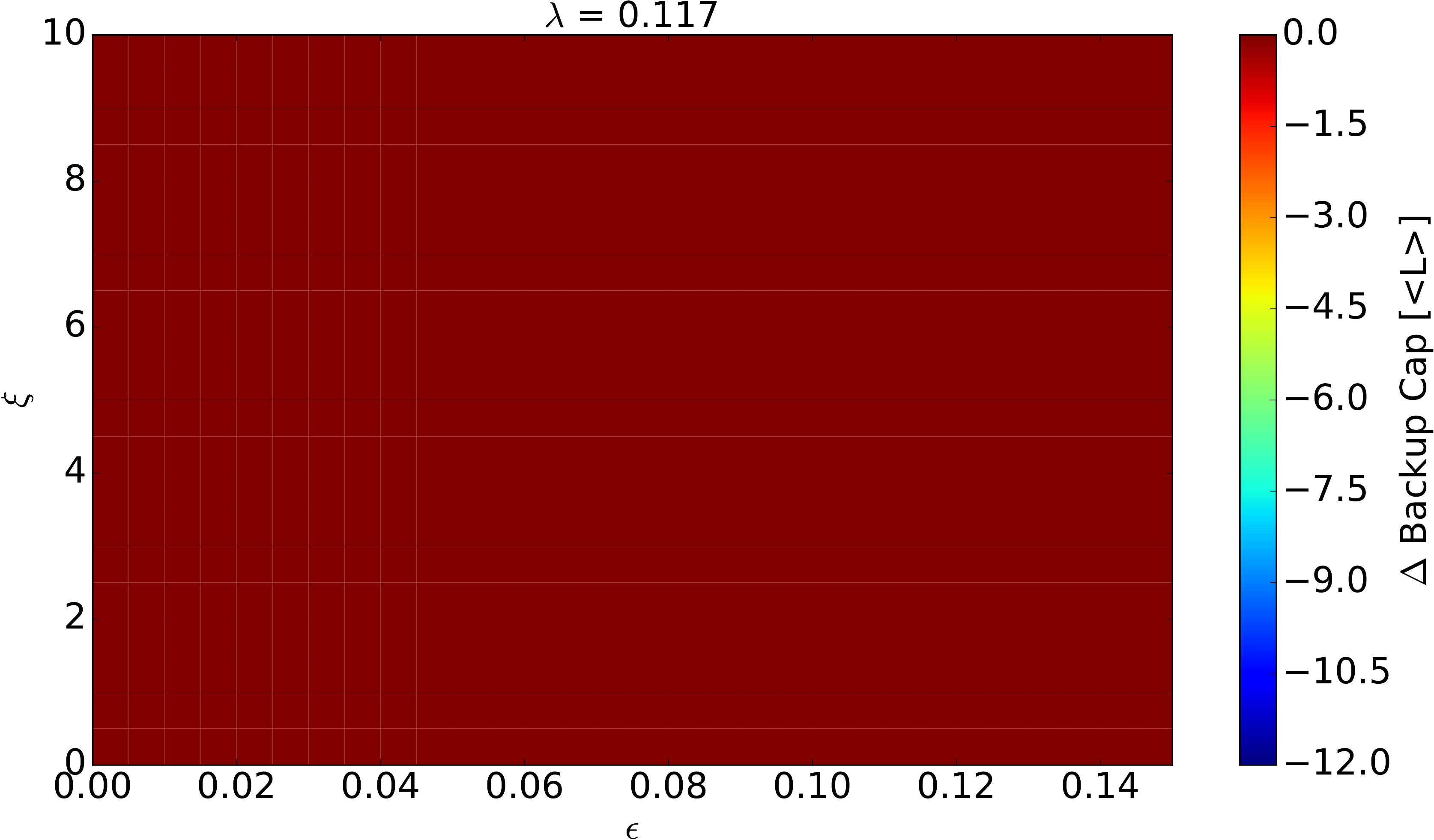}\\
\includegraphics[width=0.5\textwidth]{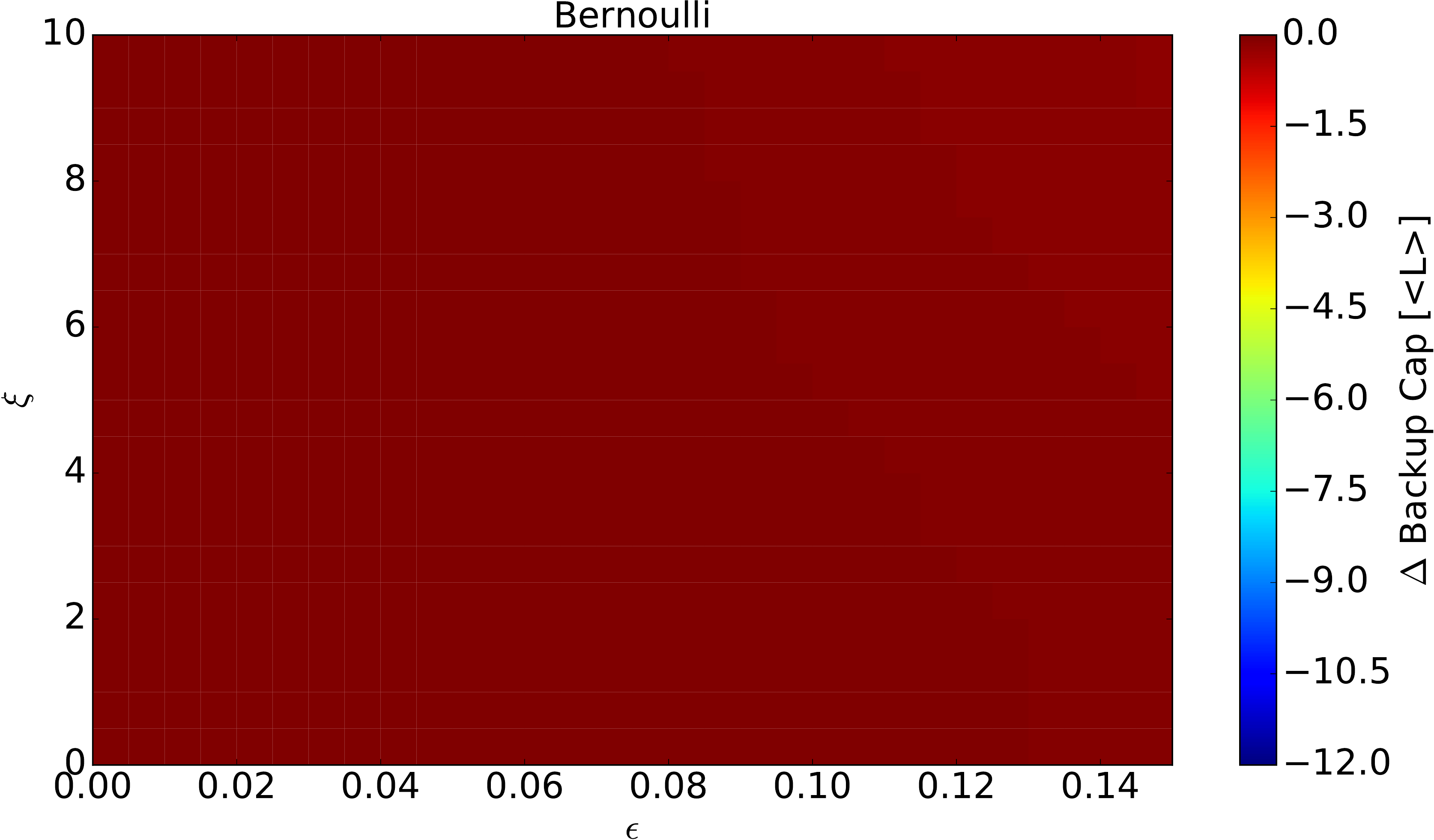}
\caption{Reduction of the European backup capacity need using North African solar imports in dependency of the average CSP generation and the CSP storage capacities for three different hydro usage strategies. Top: hydro first, middle: threshold, bottom: Bernoulli. For details, we refer to the text.}
\label{fig:backupebackupc_contour}
\end{figure}

 In addition to hydro power, CSP imports from North Africa are allowed throughout this section.
 Thus, first the mismatch between the generation from wind/PV and the load is calculated. Second, hydro power is used according to the corresponding hydro usage strategy.
 Third, CSP is imported.
 Finally, the need for backup energy and capacity is determined.
We assume the combined storage size of all Northern African CSP power plants to be described by a parameter $\xi$ via
\begin{align*}
 \kappa^S_{\text{NA}} &= \xi \cdot \left<G^{CSP}_{\text{NA}}\right>,
\end{align*}
where $\left<G^{CSP}_{\text{NA}}\right>$ is the average hourly dispatch from North African CSP.
In addition to the combined storage size, the overall generation from CSP in Northern Africa is described by the parameter $\epsilon$ via
\begin{align*}
 \left<G_{\text{NA}}^{CSP}\right> &= \epsilon \left<L\right>,
\end{align*}
and varied throughout the results.
In addition, we assume a constant solar field multiple of 2, i.e. the generation capacity is limited at 50\% of the 
peak energy inflow. \\
 Figure \ref{fig:backupebackupc2} shows the reduction of the need for backup energy in Europe in dependency of the overall generation from CSP and the thermal storage size for the three different hydro usage strategies.
If there is no generation from CSP in North Africa, the backup energy need in a fully renewable Europe, where generation
is contributed by the sources wind/PV and hydro, is around 8\% of the average load.
This can be reduced by ca. 3.5\% through average generation from CSP reaching 14\% of the European load in combination with a storage size equivalent to
10 hours of the average CSP-generation under the ``Hydro First'' strategy.
Thus, approximately 25\% of the CSP generation could be exported to Europe.
However, this also means that 75\% of the CSP generation practically could not be integrated into the European power system.
\\For the purpose of reducing the need for backup energy further, seasonal storages or cooperation of North African CSP with European hydro power and a smart use (Fig. \ref{fig:backupebackupc2}  (top) upper right corner) of their seasonal storage capabilities would be desirable.
This can be seen in the middle and bottom figure, which show the need for backup energy for the cooperative hydro strategies.
For the Threshold strategy, $\lambda$ was set to $0.117$ and for the Bernoulli strategy to $0.0$.
It can be observed that the cooperative strategies have potential to reduce the need for backup energy.
If CSP generation totals 10\% of the European load and is combined with an 8-hour storage, the need for backup energy is reduced by 6\% using the Threshold strategy, so 60\% of the
generated CSP could be exported to Europe.
However, it can also be observed that CSP storage capacities are mandatory. 
If storage capacities are reduced to zero, the need for backup energy is reduced by only 3.4\%, so thermal storage almost doubles the potential benefit.
For the Bernoulli strategy, the reduction equals 5.2\% at $(\xi, \epsilon) = (10, 0.1)$ and 2.7\% if no storage is available.
\\ Fig. \ref{fig:backupebackupc_contour} similarly shows the need for backup capacity in dependency of $\xi$ and $\epsilon$ for the three proposed hydro usage strategies.
In the case of hydro first, backup capacities are reduced by up to 12\% of the average load. 
For the Threshold strategy, almost no capacity reduction is observed until very high values of $\epsilon$ and $\xi$. 
This is probably because the backup capacity need is already pushed down to ca. 10\% of the load and for the Bernoulli strategy at $\lambda=0$ to 40\% (Fig. \ref{fig:backupebackupc}).
\\If we define a required transmission capacity $T^C$ equivalently to the required backup capacity (Eq. \ref{eq:backupcap})
via
 \begin{align}
 0.99 = \int_0^{T^C} p(I(t)) dp,
\end{align}
we find it to be ca. 70 GW of required transmission capacity for $\epsilon=0.1, \xi = 8$.

\section{Summary and Conclusion}
In this paper, we have investigated the interplay of European hydro power and North African solar thermal power in a highly renewable European power system.
We have investigated 
different hydro usage strategies and shown that a hydro first strategy
is not optimal with respect to the minimisation of the need for backup capacity and, if coupled to the North African solar power, the need for backup energy, as well.
This hydro first strategies uses hydro to cover residual mismatches after generation from the renewable sources of wind and PV without consideration of possible CSP imports.
\\In the case of Europe alone, using the hydro first strategy requires large amounts of backup capacity at around 70\% of the average hourly load, which equals around 280 GW using the assumptions of this paper. 
However, using a different hydro strategy that can be interpreted as cooperative because it only uses hydro power for peak loads, decreases the need for backup capacity strongly.
In the case of the coupled system, it can furthermore allow for large shares of North African solar thermal power to successfully balance the European power system.
However, this requires in all investigated scenarios sufficient transmission capacities between Europe and North Africa.
We have determined them to be around 70 GW, which is ca. 50\% less than was found in different studies
(\cite{trieb2012solar}).
Besides, we assumed inter-country capacities within Europe to be unlimited.
However, European inter-country transmission capacities are still far from this point and power transmission links to Northern Africa barely exist. 
Therefore, it would require enormous investments into the grid.
We conclude that theoretically European hydro power can be well combined with North African CSP to balance a highly renewable European power system, if an appropriate seasonal hydro usage strategy is chosen and the infrastructure suffices.
\section*{Acknowledgments}
The work is part of the RESTORE 2050 project (Wuppertal Institute, Next Energy, ForWind, University of Oldenburg) that
is financed by the Federal Ministry of Education and Research (BMBF, Fkz. 03SFF0439A). We would like to thank our project partners from Wuppertal Institute and Next Energy for helpful discussions and 
suggestions and the supply of load data.

% \fi
\IEEEpeerreviewmaketitle

% the docu ment is modified later


%\bibliographystyle{IEEEtran}
%\bibliography{IEEEabrv,./paper.bib}
%\begin{thebibliography}{1}
\bibliographystyle{unsrt}
\bibliography{kies_csp}

\begin{thebibliography}{10}

\bibitem{Eurostat_2015}
Eurostat.
\newblock Electricity generated from renewable sources.
\newblock http://epp.eurostat.ec.europa.eu, 2015.
\newblock Accessed: 2015-01-10.

\bibitem{Heide_2011}
Dominik Heide, Martin Greiner, Lueder von Bremen, and Clemens Hoffmann.
\newblock Reduced storage and balancing needs in a fully renewable european
  power system with excess wind and solar power generation.
\newblock {\em Renewable Energy}, (36):2515--2523, 2011.

\bibitem{Heide_2010}
Dominik Heide, Lueder Von~Bremen, Martin Greiner, Clemens Hoffmann, Markus
  Speckmann, and Stefan Bofinger.
\newblock Seasonal optimal mix of wind and solar power in a future, highly
  renewable europe.
\newblock {\em Renewable Energy}, 35(11):2483--2489, 2010.

\bibitem{Kies_2015}
Alexander Kies, Kabitri Nag, Lüder von Bremen, Elke Lorenz, and Detlev
  Heinemann.
\newblock Investigation of balancing effects in long term renewable energy
  feed-in with respect to the transmission grid.
\newblock {\em Adv. Sci. Res.}, 12:91--95, 2015.

\bibitem{Kies_2016wave}
Alexander Kies, Bruno~U Schyska, and Lueder von Bremen.
\newblock The optimal share of wave power in a highly renewable power system on
  the iberian peninsula.
\newblock {\em Energy Reports}, 2:221--228, 2016.

\bibitem{franccois2016increasing}
B~Fran{\c{c}}ois, B~Hingray, D~Raynaud, M~Borga, and JD~Creutin.
\newblock Increasing climate-related-energy penetration by integrating
  run-of-the river hydropower to wind/solar mix.
\newblock {\em Renewable Energy}, 87:686--696, 2016.

\bibitem{jurasz2017integrating}
Jakub Jurasz and Bart{\l}omiej Ciapa{\l}a.
\newblock Integrating photovoltaics into energy systems by using a
  run-off-river power plant with pondage to smooth energy exchange with the
  power gird.
\newblock {\em Applied Energy}, 198:21--35, 2017.

\bibitem{jurasz2017modeling}
Jakub Jurasz.
\newblock Modeling and forecasting energy flow between national power grid and
  a solar--wind--pumped-hydroelectricity (pv--wt--psh) energy source.
\newblock {\em Energy Conversion and Management}, 136:382--394, 2017.

\bibitem{Rasmussen_2012}
Morten~Grud Rasmussen, Gorm~Bruun Andresen, and Martin Greiner.
\newblock Storage and balancing synergies in a fully or highly renewable
  pan-european power system.
\newblock {\em Energy Policy}, 51:642--651, 2012.

\bibitem{Weitemeyer_2015}
Stefan Weitemeyer, David Kleinhans, Lukas Wienholt, Thomas Vogt, and Carsten
  Agert.
\newblock A european perspective: potential of grid and storage for balancing
  renewable power systems.
\newblock {\em Energy Technology}, 4(1):114--122, 2016.

\bibitem{weitemeyer2015european}
Stefan Weitemeyer, David Kleinhans, Lukas Wienholt, Thomas Vogt, and Carsten
  Agert.
\newblock A european perspective: potential of grid and storage for balancing
  renewable power systems.
\newblock {\em Energy Technology}, 4(1):114--122, 2016.

\bibitem{schlachtberger2016backup}
DP~Schlachtberger, S~Becker, S~Schramm, and M~Greiner.
\newblock Backup flexibility classes in emerging large-scale renewable
  electricity systems.
\newblock {\em Energy Conversion and Management}, 2016.

\bibitem{Becker_2014}
Sarah Becker, Rolando~A Rodriguez, Gorm~B Andresen, Stefan Schramm, and Martin
  Greiner.
\newblock Transmission grid extensions during the build-up of a fully renewable
  pan-european electricity supply.
\newblock {\em Energy}, 64:404--418, 2014.

\bibitem{Steinke_2010}
F.~Steinke, P.~Wolfrum, and C.~Hoffmann.
\newblock Grid vs. storage in a 100\% renewable europe.
\newblock {\em Renewable Energy}, 50:826--832, 2013.

\bibitem{kies2016curtailment}
A~Kies, BU~Schyska, and L~von Bremen.
\newblock Curtailment in a highly renewable power system and its effect on
  capacity factors.
\newblock {\em Energies}, 9(7):510, 2016.

\bibitem{kleinhans2014towards}
David Kleinhans.
\newblock Towards a systematic characterization of the potential of demand side
  management.
\newblock {\em arXiv preprint arXiv:1401.4121}, 2014.

\bibitem{Kies_2016dsm}
Alexander Kies, Bruno~U Schyska, and Lueder von Bremen.
\newblock The demand side management potential to balance a highly renewable
  european power system.
\newblock {\em Energies}, 9(11):955, 2016.

\bibitem{hirth2016system}
Lion Hirth and Simon M{\"u}ller.
\newblock System-friendly wind power: How advanced wind turbine design can
  increase the economic value of electricity generated through wind power.
\newblock {\em Energy Economics}, 56:51--63, 2016.

\bibitem{chattopadhyay2017impact}
Kabitri Chattopadhyay, Alexander Kies, Elke Lorenz, L{\"u}der von Bremen, and
  Detlev Heinemann.
\newblock The impact of different pv module configurations on storage and
  additional balancing needs for a fully renewable european power system.
\newblock {\em Renewable Energy}, 113:176--189, (2017).

\bibitem{FraunhoferISE_2013}
Fraunhofer~Institute for Solar Energy Systems~ISE.
\newblock Levelized cost of electricity renewable energy technologies, 2013.

\bibitem{trieb2012solar}
Franz Trieb, Christoph Schillings, Thomas Pregger, and Marlene O'Sullivan.
\newblock Solar electricity imports from the middle east and north africa to
  europe.
\newblock {\em Energy Policy}, 42:341--353, 2012.

\bibitem{Viebahn_2011}
P.~Viebahn, Y.~Lechon, and F.~Trieb.
\newblock The potential role of concentrated solar power (csp) in africa and
  europe—a dynamic assessment of technology development, cost development and
  life cycle inventories until 2050.
\newblock {\em Energy Policy}, 39(8):4420--4430, 2011.

\bibitem{trieb2006trans}
Franz Trieb, C~Schillings, S~Kronshage, P~Viebahn, M~Kabariti, K~Daoud,
  A~Jordan, A~Bennouna, H~el~Nokrashy, S~Hassan, et~al.
\newblock Trans-mediterranean interconnection for concentrating solar power.
\newblock {\em Forschungsvorhaben im Auftrag des Auftrag des Bundesministeriums
  f{\"u}r Umwelt, Naturschutz und Reaktorsicherheit (BMU)}, 2006.

\bibitem{Guerrero_2013}
R.~Guerrero-Lemus and J.M. Martinez-Duart.
\newblock {\em Renewable Energies and CO2}.
\newblock Springer, 3 edition, 2013.

\bibitem{pfluger2011tangible}
Benjamin Pfluger, Frank Sensfu{\ss}, Gerda Schubert, and Johannes Leisentritt.
\newblock Tangible ways towards climate protection in the european union (eu
  long-term scenarios 2050).
\newblock {\em Fraunhofer ISI}, 2011.

\bibitem{rienecker2011merra}
Michele~M Rienecker, Max~J Suarez, Ronald Gelaro, Ricardo Todling, Julio
  Bacmeister, Emily Liu, Michael~G Bosilovich, Siegfried~D Schubert, Lawrence
  Takacs, Gi-Kong Kim, et~al.
\newblock Merra: Nasa's modern-era retrospective analysis for research and
  applications.
\newblock {\em Journal of Climate}, 24(14):3624--3648, 2011.

\bibitem{Cano_1986}
Daniel Cano, Jean-Marie Monget, Michel Albuisson, Herv{\'e} Guillard, Nathalie
  Regas, and Lucien Wald.
\newblock A method for the determination of the global solar radiation from
  meteorological satellite data.
\newblock {\em Solar Energy}, 37(1):31--39, 1986.

\bibitem{Hammer_1998}
A~Hammer, D~Heinemann, A~Westerhellweg, Pierre Ineichen, JA~Olseth,
  A~Skartveit, D~Dumortier, M~Fontoynont, L~Wald, HG~Beyer, et~al.
\newblock Derivation of daylight and solar irradiance data from satellite
  observations.
\newblock In {\em Proceedings 9th Conference on Satellite Mete orology and
  Oceanography}, pages 747--750, 1998.

\bibitem{Klucher_1979}
T.M. Klucher.
\newblock Evaluation of models to predict insolation on tilted surfaces.
\newblock {\em Solar Energy}, 23(2):111--114, 1979.

\bibitem{dee2011era}
DP~Dee, SM~Uppala, AJ~Simmons, Paul Berrisford, P~Poli, S~Kobayashi, U~Andrae,
  MA~Balmaseda, G~Balsamo, P~Bauer, et~al.
\newblock The era-interim reanalysis: Configuration and performance of the data
  assimilation system.
\newblock {\em Quarterly Journal of the Royal Meteorological Society},
  137(656):553--597, 2011.

\bibitem{SAM_NREL}
M.J. Wagner and P.~Gilman.
\newblock Technical manual for the sam physical trough model.
\newblock Technical report, NREL/TP-5500-51825, 1617 Cole Boulevard Golden,
  Colorado 80401, June 2011.

\bibitem{restorereport2016}
Alexander Kies, Kabitri Chattopadhyay, Lueder von Bremen, Elke Lorenz, and
  Detlev Heinemann.
\newblock Simulation of renewable feed-in for power system studies.
\newblock Technical report, RESTORE 2050 project report, 2016.

\end{thebibliography}


\end{document}